# Modeling of SCADA and PMU Measurement Chains

Gang Cheng[1], *Graduate Student Member, IEEE* and Yuzhang Lin[2], *Member, IEEE*

[1]Department of electrical and computer engineering, University of Massachusetts Lowell, USA
[2]Department of electrical and computer engineering, New York University, USA
[1]gang_cheng@student.uml.edu; [2]yuzhang.lin@nyu.edu

In this document, the supervisory control and data acquisition (SCADA) and phasor measurement unit (PMU) measurement chain modeling will be studied, where the measurement error sources of each component in the SCADA and PMU measurement chains and the reasons leading to measurement errors exhibiting *non-zero-mean*, *non-Gaussian*, and *time-varying* statistical characteristic are summarized and analyzed. This document provides a few equations, figures, and discussions about the details of the SCADA and PMU measurement error chain modeling, which are intended to facilitate the understanding of how the measurement errors are designed for each component in the SCADA and PMU measurement chains. The measurement chain models described here are also used for synthesizing measurement errors with realistic characteristics in simulation cases to test the developed algorithms or methodologies.

## A. SCADA Measurement Chain Modeling

For the SCADA measurement chain, four major components, including *instrument transformers*, *control cables and burdens*, *intelligent electronic devices (IEDs)*, and *communication networks (CNs)*, are investigated in this document [1]. The diagram of the developed SCADA measurement chain modeling is shown in Fig. 1, and the errors introduced by each component are elaborated as follows.

*1) Instrument Transformers*. The IEEE Std C57.13-2016 has demonstrated that instrument transformers may introduce both *ratio* and *phase angle errors* [2]. In this document, it is assumed that both *ratio* and *phase angle errors* introduced by voltage transformers (VTs) and current transformers (CTs) consist of *systematic* and *random* errors. i) As discussed in the IEEE Std C57.13-2016, *systematic errors* are typically introduced by VTs and CTs since they may NOT be fully calibrated in the real-world power system [2]. In this document, it is assumed that *systematic errors* are determined by the ratio correction factor (RCF) and phase angle errors [2]-[3]. Specifically, for VTs and CTs, *systematic errors* must satisfy the accuracy requirement when instrument transformers are produced by manufacturers. The accuracy requirement of *ratio* and *phase angle errors* for VTs and CTs are presented in Figs. 2 and 3, respectively. In other words, for different accuracy classes, *systematic errors* of *ratios* and *phase angles* must be located within the corresponding parallelogram areas. For example, for VTs with the 0.6 accuracy class, the *systematic errors* should be located within the middle parallelogram area in Fig. 2. ii) Since there is no literature or standard that has thoroughly studied the *random errors* for instrument transformers. Hence, in this document, the *random errors* of *ratios* and *phase angles* are assumed to follow *non-Gaussian* distributions and synthesized by using the Gaussian mixture model (GMM). To allow user-defined properties of the GMM distributions, a parameter determination procedure is proposed. *It allows users to customize the GMM distribution and determine the similarity between the designed GMM distribution and a Gaussian distribution*. The flowchart of the proposed procedure is shown in Fig. 4, where $K$ represents the number of GMM components; $\sigma^{GAU}$ is the total standard deviation of *random errors*; $\mu^{GAU}$ is the total mean of *random errors*; $N$ is the number of samples; $\hat{\boldsymbol{q}}^{GMM}$, $\hat{\boldsymbol{\sigma}}^{GMM}$, and $\hat{\boldsymbol{\mu}}^{GMM}$ are the estimated weight, standard deviation, and mean vectors of the GMM distribution, respectively; $\hat{\bar{\sigma}}^{GMM}$ is the total standard deviation of the estimated GMM distribution, and $\eta$ and Kullback-Leibler divergence (KLD) are the pre-set and calculated similarity indices, respectively. In this document, the KLD is adopted to quantify the similarity between the GMM distribution and the Gaussian distribution with parameters $\sigma^{GAU}$ and $\mu^{GAU}$. When generating the parameters $\hat{\boldsymbol{q}}^{GMM}$ and $\hat{\boldsymbol{\mu}}^{GMM}$, only the parameters of $K$-1 GMM components need to be determined since that of the $K$-th component can be calculated based on the parameters of $K$-1 components. When generating parameters $\hat{\boldsymbol{\sigma}}^{GMM}$, they are randomly selected for all GMM components. In this procedure, the random search algorithm is used to find the feasible solution satisfying the similarity index and the total standard deviation requirement. Taking these errors into account, the phasor measurement data output from VTs and CTs can be expressed as follows.

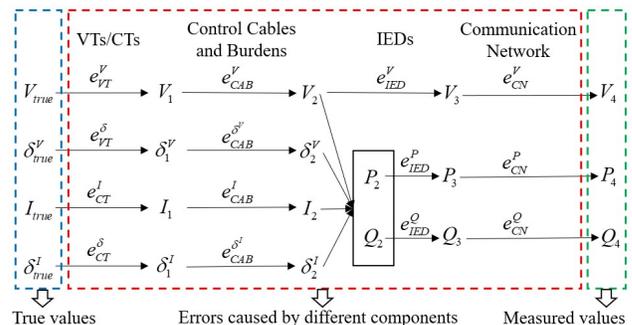

Fig. 1.  SCADA measurement chain modeling.

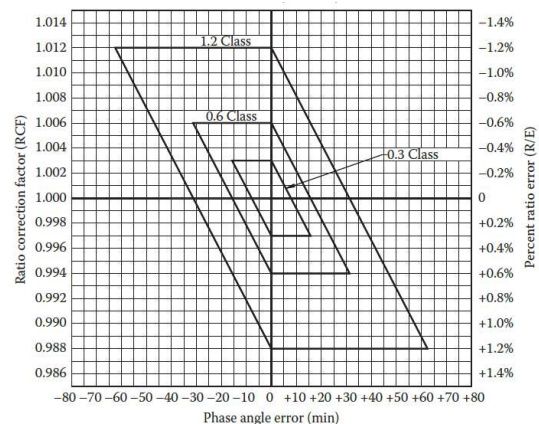

Fig. 2.  Accuracy coordinates for VTs [3].

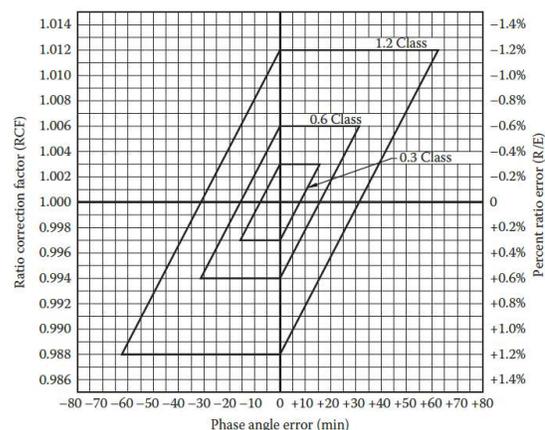

Fig. 3.  Accuracy coordinates for CTs (100% rated current) [3].

- Voltage Phasor Measurements

$$V_1 = V_{true} \cdot \left(1 + e_{VT}^V\right), \tag{1}$$

$$\delta_1^V = \delta_{true}^V + e_{VT}^\delta, \tag{2}$$

where $V_{true}$ and $\delta_{true}^V$ represent the true voltage magnitude and phase angle measurements, respectively; $V_1$ and $\delta_1^V$ represent voltage magnitude and phase angle measurements output from VTs, respectively; $e_{VT}^V$ and $e_{VT}^\delta$ are voltage magnitude and phase angle measurement errors introduced by VTs, respectively.

- Current Phasor Measurements

$$I_1 = I_{true} \cdot \left(1 + e_{CT}^I\right), \tag{3}$$

$$\delta_1^I = \delta_{true}^I + e_{CT}^\delta, \tag{4}$$

where $I_{true}$ and $\delta_{true}^I$ represent the true current magnitude and phase angle measurements, respectively; $I_1$ and $\delta_1^I$ represent current magnitude and phase angle measurements output from CTs, respectively; and $e_{CT}^I$ and $e_{CT}^\delta$ are current magnitude and phase angle measurement errors introduced by CTs, respectively.

*2) Control Cables and Burdens*. In existing published literature, it has been illustrated that control cables will introduce a time delay, which will be transformed into phase angle errors [4]. Moreover, burden resistances may also result in systematic errors for both *ratio* and *phase angle* measurements [5]. To syn-thesize these measurement errors introduced by both control cables and burdens, the *non-zero-mean* Gaussian distributions are adopted in this document. Hence, considering errors introduced by control cables and burdens, phasor measurements can be expressed as follows.

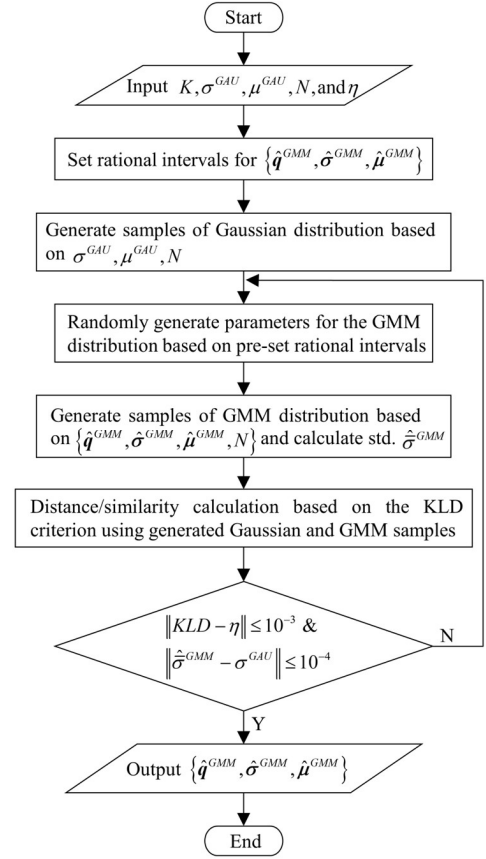

Fig. 4. Flowchart of the proposed GMM parameter estimation procedure.

$$V_2 = V_1 + e_{CAB}^V, \tag{5}$$

$$\delta_2^V = \delta_1^V + e_{CAB}^\delta, \tag{6}$$

$$I_2 = I_1 + e_{CAB}^I, \tag{7}$$

$$\delta_2^I = \delta_1^I + e_{CAB}^\delta, \tag{8}$$

where $V_2$ ($\delta_2^V$) and $I_2$ ($\delta_2^I$) represent voltage magnitude (phase angle) and current magnitude (phase angle) output from the control cables and burdens, respectively; $e_{CAB}^V$ ($e_{CAB}^\delta$) and $e_{CAB}^I$ ($e_{CAB}^\delta$) represent the magnitude (phase angle) errors of voltage and current phasor measurements, respectively.

*3) IEDs*. The industry standard definition of an IED is "any devices incorporating one or more processors with the capability to receive or send data/control from or to an external source" [1]. Based on Eqs. (10) and (12), it can be demonstrated that the power calculation within IEDs will fuse measurement errors, further deviating them from Gaussian distributions even if the original errors follow a Gaussian distribution. In addition, the IEDs may also introduce unknown random noise when they conduct A/D or D/A conversions [1]. Hence, in this document, it is assumed that IEDs will introduce random errors, which are synthesized by using *zero-mean* Gaussian distributions. The measurement data output from IEDs can be expressed as follows.

$$V_3 = V_2 + e_{IED}^V, \tag{9}$$

$$P_2 = V_2 \cdot I_2 \cdot \cos\left(\delta_2^V - \delta_2^I\right), \tag{10}$$

$$P_3 = P_2 + e_{IED}^P, \tag{11}$$

$$Q_2 = V_2 \cdot I_2 \cdot \sin\left(\delta_2^V - \delta_2^I\right), \tag{12}$$

$$Q_3 = Q_2 + e_{IED}^Q, \tag{13}$$

where $P_2$ and $Q_2$ represent the active power and reactive power calculated within IEDs, respectively; $V_3$, $P_3$, and $Q_3$ represent voltage magnitude, active power, and reactive power measurements output from IEDs, respectively; and $e_{IED}^V$, $e_{IED}^P$, and $e_{IED}^Q$ represent voltage magnitude, active power, and reactive power measurement errors introduced by IEDs, respectively.

*4) Communication Networks*. Two important characteristics of digitized measurements need to be considered when they are transmitted via CNs in SCADA systems [1]: *latency* and *time skew*. Moreover, received measurement data may not be immediately used to perform static state estimation (SSE) in control centers since SSE is conducted at a constant frequency. Hence, *buffer* time also needs to be considered in power system state estimation. The brief descriptions of *latency*, *time skew*, and *buffer* are as follows:

- *Latency* is the time delay during which measured data is transmitted from IEDs to the control center via CNs, which is impacted by bandwidths, transmission rates, congestion, etc.
- *Time skew* is the time difference between two measurement scans in a data set, which is introduced by the asynchrony of SCADA system measurements.
- *Buffer* is the time from receiving the measurement set to their use for power system state estimation.

In realistic power systems, the total time delay will introduce measurement errors, which is shown in Fig. 5, with the fluctuation of operating states. These errors are typically unpredictable and complex [6]-[7]. To model the measurement errors introduced by CNs, two different schemes are designed in this report, *i.e.*, *Real-world-data-based error modeling* and *GMM-based error modeling*.

**Scheme 1: Real-World-Data-Based Error Modeling**

In Scheme 1, measurement errors introduced by CNs, i.e., the total time delay and fluctuations of measurement values, are extracted from the realistic measurement data. Hence, this scheme requires abundant historical measurement data so that users can extract measurement errors with respect to different total time delays and then model the distribution of them.

***First, samples of total time delays need to be generated.*** Total time delay consists of *latency* and *buffer*. *Latency* is impacted by different factors, such as transmission rate, package length, congestion, etc. Typically, CN *latency* is statistically modeled using Weibull, Lognormal, Gamma, Exponential distributions, and their hybrids [8]. In this document, CN *latency* is modeled using the Lognormal mixture model (LMM) distribution.

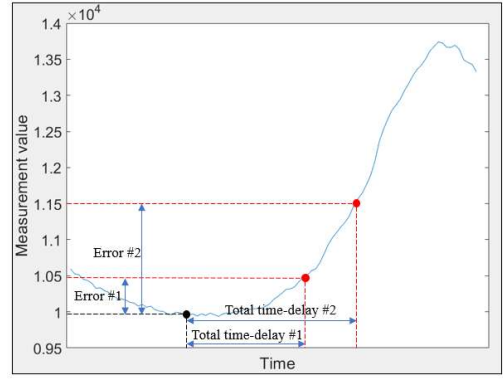

Fig. 5. Measurement errors introduced by communication networks in the SCADA measurement chain.

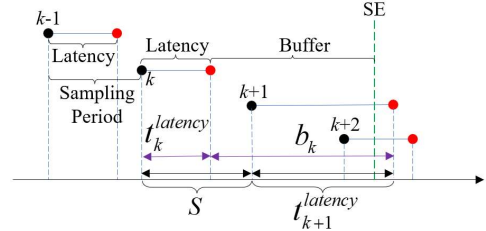

Fig. 6. Latency is smaller than the sampling period.

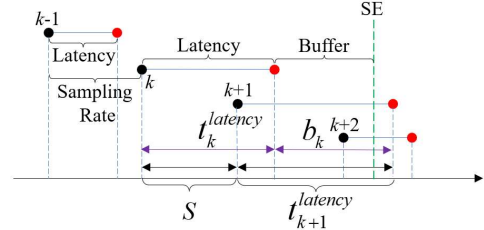

Fig. 7. Latency is larger than the sampling period.

$$L(x) = \sum_{i=1}^{K} \frac{q_i}{x\sigma_i\sqrt{2\pi}} \cdot \exp\left(-\frac{(\ln x - \mu_i)^2}{2\sigma_i^2}\right), \tag{14}$$

where $x$ is the *latency* variable; $K$ is the number of LMM components, and $q_i$, $\mu_i$, and $\sigma_i$ are the weight, mean, and standard deviation of the $i$th LMM component, respectively.

To create samples of CN *latency*, users need to set the values for the parameters of the LMM distribution. Then, the MATLAB function "*lognrnd*" can be adopted to generate samples that follow the designed LMM distribution.

*Buffer* is impacted by *latency* and the *sampling period*. The relationship among *latency*, *sampling period*, and *buffer* is shown in Figs. 6 and 7. In these figures, two situations are discussed, i.e., *latency* is smaller (Fig. 6) or larger (Fig. 7) than the *sampling period*. Based on this figure, the limits of *buffer* can be derived as follows,

$$b_k = S + t_{k+1}^{latency} - t_k^{latency}, \tag{15}$$

where $b_k$ is represents the limit of the *buffer* of the measurement at the $k$th time instant; $S$ is the sampling period; $t_k^{latency}$ is the *latency* of the measurement sent by IEDs at the $k$th time instant; and $k$ is a positive integer. In the following contents, the measurement sent by IEDs at the $k$th time instant will be referred to as the $k$th sample for simplicity. If $b_k$ is less than zero, then the $k$th sample will be discarded when conducting SSE since the ($k$+1)th sample will arrive at the control center earlier than the $k$th one. Hence, the ($k$+1)th measurement data will be used to perform SSE.

Based on the above analysis, the distributions of *latency* and *buffer* are as follows,

- Latency - it is assumed that *latency* follows a LMM distribution,

$$t_k^{latency} \sim LMM\left(\phi^{latency}\right) \quad (16)$$

- Buffer - it is assumed that the *buffer* follows a **uniform distribution**,

$$t_k^{buffer} \sim U\left(0, b_k\right), \quad (17)$$

where $\phi^{latency}$ represents the parameter set of the LMM distribution, including weights, mean, and standard deviations in Eq. 14.

Hence, the *total time delay* can be expressed as follows:

$$t_k^{delay} = t_k^{latency} + t_k^{buffer}. \quad (18)$$

***Second, measurement errors introduced by CNs, i.e., the total time delay and fluctuations of measurement values, are extracted from the real-world measurement data***. The derived measurement errors are as follows,

$$e_{CN}^V = z^V\left(t_k + t_k^{delay}\right) - z^V\left(t_k\right) \quad (19)$$

$$e_{CN}^P = z^P\left(t_k + t_k^{delay}\right) - z^P\left(t_k\right) \quad (20)$$

$$e_{CN}^Q = z^Q\left(t_k + t_k^{delay}\right) - z^Q\left(t_k\right) \quad (21)$$

where $t_k$ represents the $k$th time instant; $t_k^{delay}$ is the total time delay of the $k$th sample; $z^V$, $z^P$, and $z^Q$ represent the voltage magnitude, active power, and reactive power measurements, respectively; $z\left(t_k\right)$ represent the measurement at $t_k$; and $e_{CN}^V$, $e_{CN}^P$, and $e_{CN}^Q$ represent the voltage magnitude, active power, and reactive power measurement errors introduced by CNs, respectively.

**Scheme 2: GMM-Based Error Modeling**

In Scheme 1, it has been demonstrated that the measurement errors introduced by CNs may follow a non-Gaussian distribution by exploring the real-world datasets (EPFL-campus PMU dataset [9]). Hence, in Scheme 2, a GMM-based error modeling method is proposed. Similar to the *random error* modeling in instrument transformers, GMM is also adopted in this subsection to model the *random error* introduced by CNs. The proposed GMM distribution of *random errors* is as follows,

$$e_{CN}^V \sim GMM\left(\phi_{CN}^V\right) \quad (22)$$

$$e_{CN}^P \sim GMM\left(\phi_{CN}^P\right) \quad (23)$$

$$e_{CN}^Q \sim GMM\left(\phi_{CN}^Q\right) \quad (24)$$

***For the determination of the parameters of the GMM distributions, please refer to the parameter estimation procedure discussed in Fig. 4.***

In summary, the measurement error introduced by CNs can be synthesized by using **Scheme 1** or **Scheme 2**. Taking these errors into account, measurement data received in the control center can be expressed as follows.

$$V_4 = V_3 + e_{CN}^V, \quad (25)$$

$$P_4 = P_3 + e_{CN}^P, \quad (26)$$

$$Q_4 = Q_3 + e_{CN}^Q, \quad (27)$$

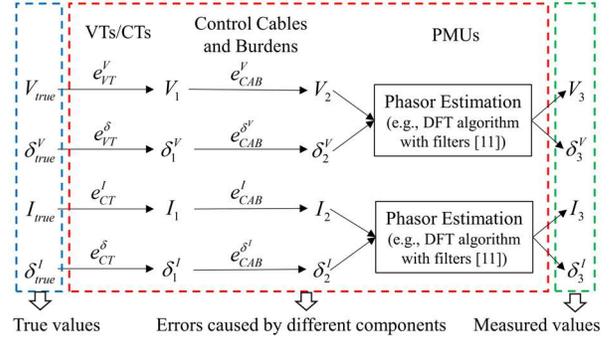

Fig. 8. PMU measurement chain modeling.

TABLE I
M CLASS LOW PASS FILTER PARAMETERS [11]

| Reporting rate $F_s$ | | Filter reference frequency $F_{fr}$ (Hz) | Filter order $N$ |
|---|---|---|---|
| 50 Hz | 10 | 1.779 | 806 |
| | 25 | 4.355 | 338 |
| | 50 | 7.75 | 142 |
| | 100 | 14.1 | 66 |
| 60 Hz | 10 | 1.78 | 968 |
| | 12 | 2.125 | 816 |
| | 15 | 2.64 | 662 |
| | 20 | 3.50 | 502 |
| | 30 | 5.02 | 306 |
| | 60 | 8.19 | 164 |
| | 120 | 16.25 | 70 |

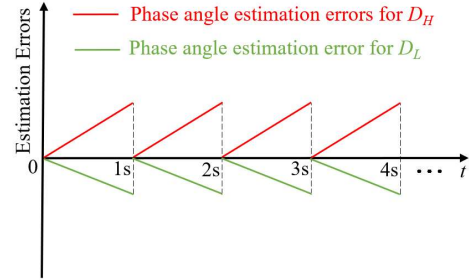

Fig. 9. Phase angle estimation errors introduced by sampling time errors in PMUs [12].

where $V_4$, $P_4$, and $Q_4$ represent voltage magnitude, active power, and reactive power measurements received in the control center, respectively; and $e_{CN}^V$, $e_{CN}^P$, and $e_{CN}^Q$ represent the voltage magnitude, active power, and reactive power errors introduced by CNs, respectively.

### B. PMU Measurement Chain Modeling

For the PMU measurement chain, three major components, including *instrument transformers*, *control cables and burdens*, and *PMUs*, are investigated in this document [10]-[14]. The diagram of the developed PMU measurement chain modeling is shown in Fig. 8, and the errors introduced by each component are elaborated as follows.

*1) Instrument Transformers*. The measurement errors introduced by VTs and CTs in the PMU measurement chain are the same as those in the SCADA measurement chain.

*2) Control cables and Burdens*. The measurement errors introduced by control cables and burdens in the PMU measurement chain are the same as those in the SCADA measurement chain.

*3) PMUs.* In synchronized phasor measurements, it has been studied that phasor estimation is conducted within PMUs by using the sampled values collected from instrument transformers. Various phasor estimation algorithms have been proposed, such as discrete Fourier transform (DFT) methods, least squares methods, Kalman filter methods, and Prony methods [15]. In this document, the DFT algorithm with low pass filters is adopted to implement phasor estimation [11], which is presented as follows.

Given a set of sampled values of the input signal, the synchrophasor estimates at the time of the *i*th sampled value can be expressed by,

$$\hat{V}_3(i) = \frac{\sqrt{2}}{G} \times \sum_{k=-N/2}^{N/2} V_2^{SV}(i+k) \times W(k) \times e^{-j(i+k)\cdot\Delta t\cdot\omega_0}, \quad (28)$$

$$V_2^{SV} = V_2 \cdot \cos\left(2\pi \cdot f_0 \cdot t + \delta_2^V\right), \quad (29)$$

$$\hat{I}_3(i) = \frac{\sqrt{2}}{G} \times \sum_{k=-N/2}^{N/2} I_2^{SV}(i+k) \times W(k) \times e^{-j(i+k)\cdot\Delta t\cdot\omega_0}, \quad (30)$$

$$I_2^{SV} = I_2 \cdot \cos\left(2\pi \cdot f_0 \cdot t + \delta_2^I\right), \quad (31)$$

where $V_2^{SV}(i+k)$ and $I_2^{SV}(i+k)$ represent the sampled voltage and current waveform data at the time of the $(i+k)$th sampled values, respectively; $\omega_0$ is the angular frequency $2\pi f_0$, where $f_0$ is the nominal power system frequency (50 Hz or 60 Hz); $N$ is the finite impulse response (FIR) filter order, which is shown in Table I; $\Delta t$ is the sampling interval, which is equal to $1/F_{sampling}$, where $F_{sampling}$ is the sampling frequency, which is 960 Hz for the 60 Hz reference model [11]; $G$ is the gain, which is equal to $\sum_{k=-N/2}^{N/2} W(k)$; and $W(k)$ is the low pass filter coefficient, which is given as follows:

$$W(k) = \frac{\sin\left(2\pi \times \frac{2F_{fr}}{F_{sampling}} \times k\right)}{2\pi \times \frac{2F_{fr}}{F_{sampling}} \times k} h(k), \quad (32)$$

$$h(k) = 0.54 + 0.46 \times \cos\left(\frac{2\pi \cdot k}{N}\right), \quad (33)$$

where $F_{fr}$ is the low pass filter reference frequency, which is shown in Table I; $h(k)$ is the hamming function; and $W(0)$ is defined as 1 since $W=0/0$ is indeterminate.

In PMUs, it has been illustrated that sampling time errors [12], off-nominal frequency of input signals [13], and global positing system (GPS) signal loss [14] can lead to phasor estimation errors. In this document, the following contents will provide detailed discussions about these factors.

• *Sampling time error*

Sampling time errors are typically introduced by the frequency drift of oscillators and the fact that the sampling clock is not precisely at a multiple of the power system frequency [12], which will make the sampling time of the *i*th sampled value inaccurate. For example, the accurate sampling time of the *i*th sampled value should be $t = i \cdot \Delta t$, but the actual sampling time may be $i \cdot \Delta t + t_{error}$, thereby resulting in phase angle estimation errors, which can be expressed as follows,

$$\varepsilon_\delta = 360^\circ \times t_{error} \times f_0, \quad (34)$$

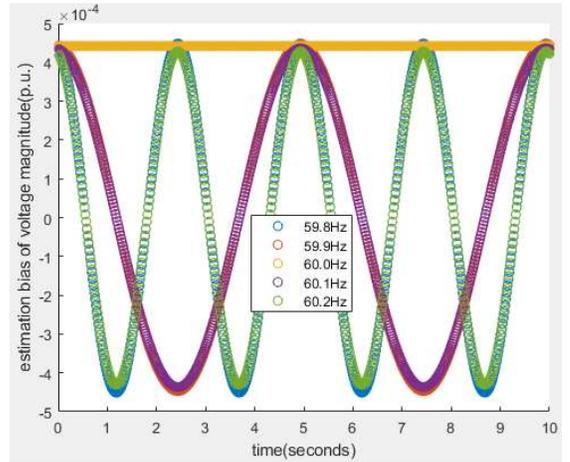

Fig. 10. Voltage magnitude estimation error introduced by off-nominal frequency in PMUs [13].

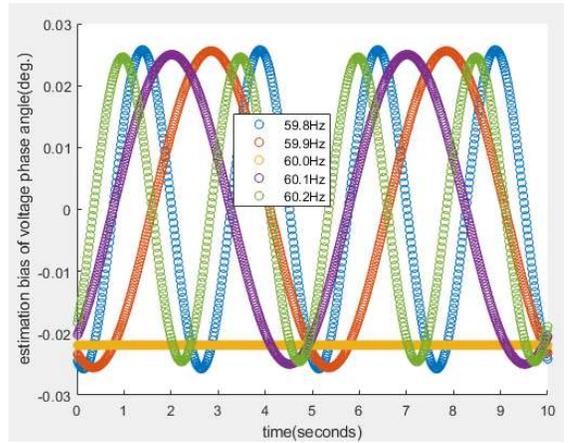

Fig. 11. Voltage phase angle estimation error introduced by off-nominal frequency in PMUs [13].

where $t_{error}$ is the sampling time error. The phase angle estimation errors are presented in Fig. 9, where $D_H$ and $D_L$ represent two different work patterns of the timer period register (PRD) [12]. It can be concluded that phase angle estimation errors will be accumulated and then cleared by the pulse per second (PPS) signal in one second, leading to phase angle measurement errors following a *non-zero-mean near-uniform* distribution.

• *Off-nominal frequency*

Off-nominal frequency is the most common phenomenon in realistic power systems due to changes in load and generation imbalances, interactions between real power demand on the network, inertia of large generators, faults, and switching events. In [13], it has been studied that the phasor estimation errors behave like a sinusoidal waveform over time. The voltage magnitude and phase angle estimation errors under different off-nominal frequencies are tested and presented in Figs. 10 and 11, respectively. Moreover, the frequency of the sinusoidal waveform can be represented by the formula as follows,

$$f_{error} = 2 \cdot |f_0 - f|, \quad (35)$$

where $f_{error}$ is the frequency of the phasor estimation errors and $f$ is the frequency of the input signal [13]. In the real world, the frequency of the input signal may be time-varying with the change of the operating state of the power system, thereby leading to phasor estimation errors following *non-Gaussian* and *time-varying* distributions.

- *GPS signal loss*

The reliability of GPS signals is one of the critical issues for synchronized measurement devices (SMDs). By demodulating the GPS signal, the GPS receiver can align its time with coordinated universal time (UTC) and then output a high-precision PPS signal for synchronization [14]. Hence, the sampling time errors can be cleared by the PPS signal in one second. Once the GPS signal is lost, SMDs should rely on their internal crystal oscillators to provide the timing reference without the GPS correction. However, crystal oscillators may have an unintentional and typically arbitrary frequency drift away from the nominal frequency, which may be caused by a few factors, such as temperature, component aging, etc. The frequency drift of the oscillators will introduce sampling time errors in SMDs [14]. Consequently, the phase angle estimation error will accumulate over the duration of the GPS signal loss, resulting in a *time-varying*, *long-tailed*, and *asymmetric* distribution.

Assuming the timing error caused by frequency drift of the oscillators is $t_{PPS}$ μs, the phase angle error affected by this timing error within one second is,

$$e^{\theta} = 2\pi \times t_{PPS} \times 10^{-6} \times f_0 \qquad (36)$$

where $e^{\theta}$ represents the phase angle error introduced by the GPS loss. For example, if the GPS signal is lost for 10 minutes and $t_{PPS}$ =0.15 μs, the time error nearly reaches 90 μs, which amounts to an angle error of 1.94° when frequency of the input signal is 60 Hz.

The average GPS loss rate for the PMU from 2009 to 2012 was 5 times per day, as reported in [16]. Moreover, the average GPS signal recovery time follows an approximate exponential distribution, as illustrated in Fig. 12 [14]. In this document, the recovery time is fitted by using an exponential distribution. By this means, the impact of GPS signal loss can be studied within a limited length of synthesized PMU data section.

The exponential distribution function is fitted as follows:

$$f(t;\lambda) = \begin{cases} 0.13e^{-0.13t}, & t \le 0 \\ 0, & t > 0 \end{cases} \qquad (37)$$

where $t$ represents the GPS signal recovery time and $\lambda$ is the parameter, which is 0.13 in Eq. 37.

In summary, the measurement error sources for each component in the SCADA and PMU measurement chains have been thoroughly examined and analyzed in this document. These theoretical analyses and examples illustrate that the measurement errors introduced by the SCADA and PMU measurement chains can indeed exhibit *non-zero-mean*, *non-Gaussian*, and *time-varying* statistical characteristics.

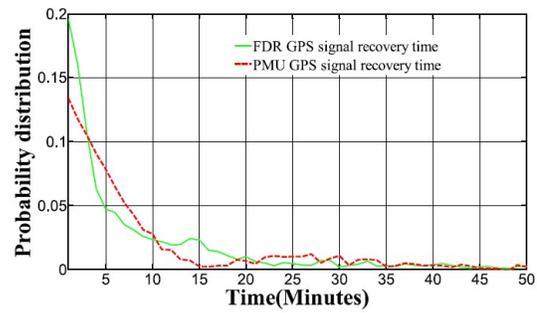

Fig. 12. Probability distribution of GPS signal recovery time for both commercial PMUs and FDRs [14].